\documentclass{article}

\usepackage{arxiv}
\usepackage[utf8]{inputenc} % allow utf-8 input
\usepackage[T1]{fontenc}    % use 8-bit T1 fonts
\usepackage{hyperref}       % hyperlinks
\usepackage{url}            % simple URL typesetting
\usepackage{booktabs}       % professional-quality tables
\usepackage{amsfonts}       % blackboard math symbols
\usepackage{nicefrac}       % compact symbols for 1/2, etc.
\usepackage{microtype}      % microtypography
\usepackage{lipsum}		% Can be removed after putting your text content
\usepackage{graphicx}
\usepackage{natbib}
\usepackage{doi}
\usepackage{amsmath}
\usepackage{float}

\title{Calculation of a force effect from muscle action to a quaternion-based musculoskeletal model}

\author{
 Ondrej Zoufaly \\
  Faculty of Mechanical Engineering\\
  Czech Technical University in Prague\\
  Prague, Technicka 1902/4, Czech Republic \\
  \texttt{ondrej.zoufaly@fs.cvut.cz} \\
   \And
 Edward Chadwick \\
  School of Engineering\\
  University of Aberdeen\\
  Aberdeen, Elphinstone Road, AB24 3UE, United Kingdom \\
  \texttt{edward.chadwick@abdn.ac.uk} \\
  \And
 Dimitra Blana \\
  School of Medicine, Medical Sciences and Nutrition\\
  University of Aberdeen\\
  Foresterhill Rd, Aberdeen AB25 2ZD, United Kingdom \\
  \texttt{dimitra.blana@abdn.ac.uk} \\
  \And
 Matej Daniel \\
  Faculty of Mechanical Engineering\\
  Czech Technical University in Prague\\
  Prague, Technicka 1902/4, Czech Republic \\
  \texttt{matej.daniel@cvut.cz} \\
}

\begin{document}
\maketitle
\begin{abstract}
 Euler angle representation in biomechanical analysis allows straightforward description of joints rotations. However, application of Euler angles could be limited due to singularity called gimbal lock. Quaternions offer an alternative way to describe rotations but they have been mostly avoided in biomechanics as they are complex and not inherently intuitive, specifically in dynamic models actuated by muscles. This study introduces a mathematical framework for describing muscle actions in dynamic quaternion-based musculoskeletal simulations. The proposed method estimates muscle torques in quaternion-based musculoskeletal model. Its application is shown on three-dimensional double-pendulum system actuated by muscle elements. Furthermore, transformation of muscle moment arms obtained from muscle paths based on Euler angles into quaternions description is presented. The proposed method is advantageous for dynamic modeling of musculoskeletal models with complex kinematics and large range of motion like the shoulder joint.
\end{abstract}

\section{Introduction}
\label{sec:Introduction}
Inter-segmental rotation is the primary type of motion observed in the human body. The standard method for describing rotations follows the notation of motions in anatomical planes.  Mathematically, it can be expressed as a sequence of rotations of generalised coordinates, also known as Euler or Tait-Bryan angles.  The rotation sequences of individual joints are provided in the ISB recommendation in \citep{wu_isb_2005}. 

In the description of systems using Euler angles, so-called gimbal lock might occur. This happens when two of the three rotational axes align, making it impossible to rotate independently around one of the axes. As a result, the description of the system loses a degree of freedom, limiting the orientations it can represent.
Gimbal lock has been identified in the shoulder joint kinematics in various activities, such as tennis \citep{bonnefoy-mazure_rotation_2010} or volleyball \citep{barrett_rotation_2024}. 
Despite different rotation sequences of Euler angles adopted for various humeral motions \citep{clark_tracking_2020, creveaux_rotation_2018,phadke_comparison_2011}, the gimbal lock cannot be avoided \citep{senk_rotation_2006}. 
Some authors even lock two axes of rotation in the glenohumeral joint to prevent gimbal lock in shoulder shrugging \citep{belli_does_2023}. Quaternion-based models would overcome problems with gimbal lock \citep{shoemake_animating_1985} and offer an approach to run upper extremity simulation without the need of changing the sequence of Euler angles based on the performed motion \citep{rab_method_2002,schnorenberg_effect_2021}.

 \citet{challis_quaternions_2020} discusses the benefits of quaternions in describing human angular kinematics. Quaternions are also widely utilized for analyzing kinematic data from IMUs \citep{cocco_comparative_2024, zhu_imu_2023} and predicting motion using neural networks \citep{cao_qmednet_2022, pavllo_modeling_2020}. However, their integration into motion analysis tools, such as OpenSim \citep{delp_opensim_2007}, remains limited. While kinematic description using quaternions is quite straightforward and well-established, the application of quaternions in dynamics has mostly been avoided in biomechanics. A comprehensive description of how muscle action contributes to the dynamical model represented by quaternions has not been provided yet. For the dynamic simulation of quaternion-based musculoskeletal models, it is essential to define the calculation of external torque resulting from muscle actions.

This study aims to present a method for describing muscle action in a dynamical musculoskeletal model using quaternions. Specifically, we provide a way to derive muscle torque in spatial coordinates. Additionally, we introduce a mapping of muscle moment arms into quaternion description and derive the associated transformations. The proposed description is verified in a three-dimensional double-pendulum model actuated by muscle elements and applied to polynomial approximation of a muscle element during humeral abduction in scapular plane.

% applied in data processing for shoulder elevation. 

\section{Methods}

\subsection{Spatial rotation velocity in terms of quaternion elements}\label{Section, eoms}

The relation between spatial rotation velocity and the quaternion time derivative can be defined as

\begin{equation}\label{eq,rotation vel from quat}
\vec{\omega} = 2\textbf{G} \vec{\dot{Q}}
\end{equation}

where $\vec{\omega}$ is the vector of spatial rotation velocities in the body reference frame and $\textbf{G}$ maps the quaternion time derivative to spatial rotation velocities \citep{graf_quaternions_2008}

\begin{equation}\label{G mapping}
\textbf{G} = \begin{bmatrix}-Q_1 & Q_0 & Q_3 & -Q_2\\\ -Q_2&-Q_3& Q_0& Q_1 \\\ -Q_3& Q_2& -Q_1& Q_0\end{bmatrix}
\end{equation}

\subsection{Derivation of external torque in quaternion coordinates}\label{Section, Map to external torque Graf}

If a muscle length can be estimated analytically as a function of quaternion elements, i.e. $l_m = f(\vec{Q})$, the muscle length Jacobian w r.t. quaternion elements $\textbf{R}_Q$ can be calculated as follows:

\begin{equation}\label{eq,jacobian in quat coord}
\textbf{R}_Q = \frac{\partial l_m(\vec{Q})}{\partial Q_i}
\end{equation}

where $\vec{Q}$ is a quaternion vector and $Q_i$ is the $i$-th element of a quaternion. The mapping between $\textbf{R}_Q$ and the muscle length Jacobian in spatial coordinates, $\textbf{R}_S$, is as follows \citep{graf_quaternions_2008}:

\begin{equation}\label{eq,jacobian in spatial coord from analytical length}
\textbf{R}_S = \frac{1}{2} \textbf{G} \textbf{R}_Q
\end{equation}

Mapping between $\textbf{R}_S$ and muscle length Jacobian in Euler coordinates (as known as moment arms) $\textbf{R}_E$ can be calculated using these relations

\begin{equation}\label{eq,Rs to moment arms}
\textbf{R}_E = \textbf{J}^T \textbf{R}_S
\end{equation}

\begin{equation}\label{eq,moment arms to RS}
\textbf{R}_S = \left( \textbf{J}^T \right)^{-1} \textbf{R}_E
\end{equation}

where $\textbf{J}$ is the system Jacobian that maps between spatial and generalized velocities.

An external (spatial) torque $\vec{\tau}$ is calculated as

\begin{equation}\label{eq,external torque from analytical length}
\vec{\tau} = \textbf{R}_S \vec{F_m}
\end{equation}

 where $\vec{F_m}$ is a vector of muscle forces.

However, mapping $\textbf{R}_S$ to the quaternion muscle length Jacobian using the following relation

\begin{equation}\label{eq,JQ from spat}
\textbf{R}_Q = 2 \textbf{G}^T \textbf{R}_S
\end{equation}

does not necessarily lead to the quaternion muscle length Jacobian calculated analytically using eq. \ref{eq,jacobian in quat coord}. This is due to the fact that the mapping in eq. \ref{eq,JQ from spat} is from 3 to 4 degrees of freedom without using any constraint, so it is not uniquely defined.

\subsection{Derivation of external torque in quaternion coordinates with additional constraint}\label{Section, mapping with cnst}

In section \ref{Section, Map to external torque Graf} it was discovered that a mapping from $\textbf{R}_S$ to the quaternion muscle length Jacobian $\textbf{R}_Q$ is ambiguous. In this approach, the constraint of the unit quaternion

\begin{equation}\label{unit quat constraint}
Q_0^2 + Q_1^2 + Q_2^2 + Q_3^2 = 1
\end{equation}

is taken and through time derivative the following constraint is obtained

\begin{equation}\label{unit quat constraint dt}
\dot{Q_0}Q_0 + \dot{Q_1}Q_1 + \dot{Q_2}Q_2 + \dot{Q_3}Q_3 = 0
\end{equation}

This constraint can be rewritten as

\begin{equation}\label{eq,constraint in matrix form}
\underbrace{\begin{bmatrix} \dot{Q_0}\\\ \dot{Q_1} \\\ \dot{Q_2} \\\ \dot{Q_3} \end{bmatrix}}_{\vec{\dot{Q}}} = 
\underbrace{\begin{bmatrix} \frac{-Q_1}{Q_0} & \frac{-Q_2}{Q_0} & \frac{-Q_3}{Q_0} \\\ 1&0&0 \\\ 0&1&0 \\\ 0&0&1 \end{bmatrix}}_\textbf{T}
\underbrace{\begin{bmatrix}  \dot{Q_1} \\\ \dot{Q_2} \\\ \dot{Q_3} \end{bmatrix}}_{\vec{\dot{Q}}^*}
\end{equation}

Substituting $\vec{\dot{Q}}$ to the eq. \ref{eq,rotation vel from quat} leads to

\begin{equation}\label{mapping between ang vel and const quat}
\vec{\omega} = \underbrace{2\textbf{G} \textbf{T}}_\textbf{E} \vec{\dot{Q}}^*
\end{equation}

where $\textbf{E}$ will be used later to map from spatial to quaternion coordinates. At this point, $Q_0$ in the analytical muscle length function needs to be substituted for $\sqrt{1-Q_1^2-Q_2^2-Q_3^2}$, so that now $l_m = f(\vec{Q}^*)$, where $\vec{Q}^* = f(Q_1, Q_2, Q_3)$. Then the muscle length Jacobian $\textbf{R}_{Q}^*$ can be calculated as follows

\begin{equation}\label{eq,RG* calculation}
\textbf{R}_{Q}^* = \frac{\partial l_m(\vec{Q}^*)}{\partial Q_i^*}
\end{equation}

The mapping between $\textbf{R}_Q^*$ and $\textbf{R}_S$ is as follows

\begin{equation}\label{eq, R_Q^* mapping}
\textbf{R}_{Q}^* = \textbf{E}^T \textbf{R}_{S}
\end{equation}

\begin{equation}\label{eq, R_S from R_Q^* mapping}
\textbf{R}_{S} = (\textbf{E}^T)^{-1} \textbf{R}_{Q}^*
\end{equation}

External torque can be estimated as follows

\begin{equation}\label{eq,external torque from analytical length const}
\vec{\tau} = \textbf{R}_S \vec{F_m}
\end{equation}

The summary of all mappings used is shown in the fig. \ref{fig,mapping_diagram}.

\begin{figure}[H]
\centering
\includegraphics[width=0.75\linewidth]{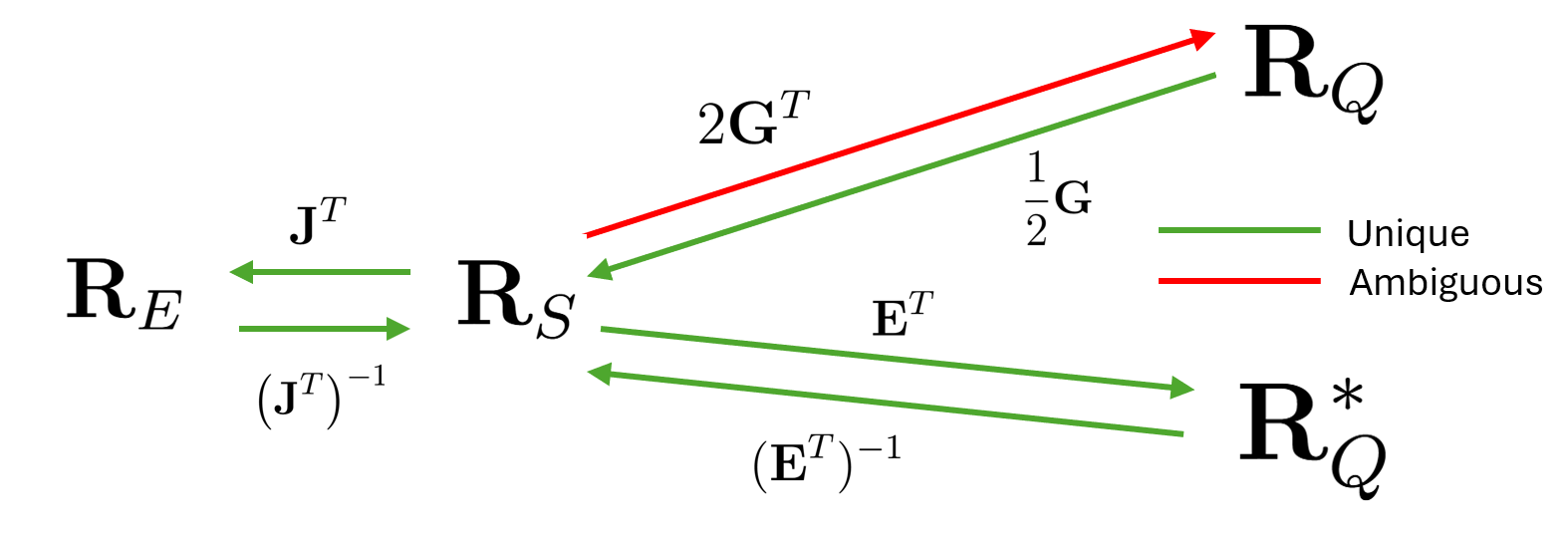}
\caption{Different mappings used in this article and their ambiguity. $\textbf{R}_{E}$ and $\textbf{R}_{S}$ are the muscle length Jacobian in Euler (as known as moment arms) and spatial coordinates, respectively. $\textbf{R}_{Q}$ and $\textbf{R}_{Q}^*$ are the unconstrained and constrained quaternion muscle length Jacobians, respectively. Note that the mapping from quaternions to spatial coordinates is unique for both approaches, however, mapping the muscle length Jacobian from spatial coordinates to quaternions uniquely is possible for constrained approach only.}
\label{fig,mapping_diagram}
\end{figure}

\subsection{Double 3D pendulum with 6 straight-line muscles}\label{Section, Pendulum description}

A simple double 3D pendulum made of 2 cylindrical rods, where each rod is of length 1 meter, radius of 0.1 meters and mass of 1 kg, was created in Matlab SimScape \citep{miller2021simscape}. Two versions of this model were created, the joint rotations in the first version are represented by YZY sequences of Euler angles and in the second version the joint rotations are represented by quaternions. The inertial properties of the rods were calculated analytically. The model includes six straight-line muscle elements represented by point-to-point forces with various attachment points. A single forward simulation was performed, during which the muscle force magnitudes were assigned randomly and kept constant. The result of the simulation consists of Euler angles and quaternions representing the rotations in individual joints. The Jacobians analysis in the result section was performed on a muscle element with an origin spanning both joints, as depicted in fig. \ref{fig,double_pend}.

\begin{figure}[H]
\centering
\includegraphics[width=0.7\linewidth]{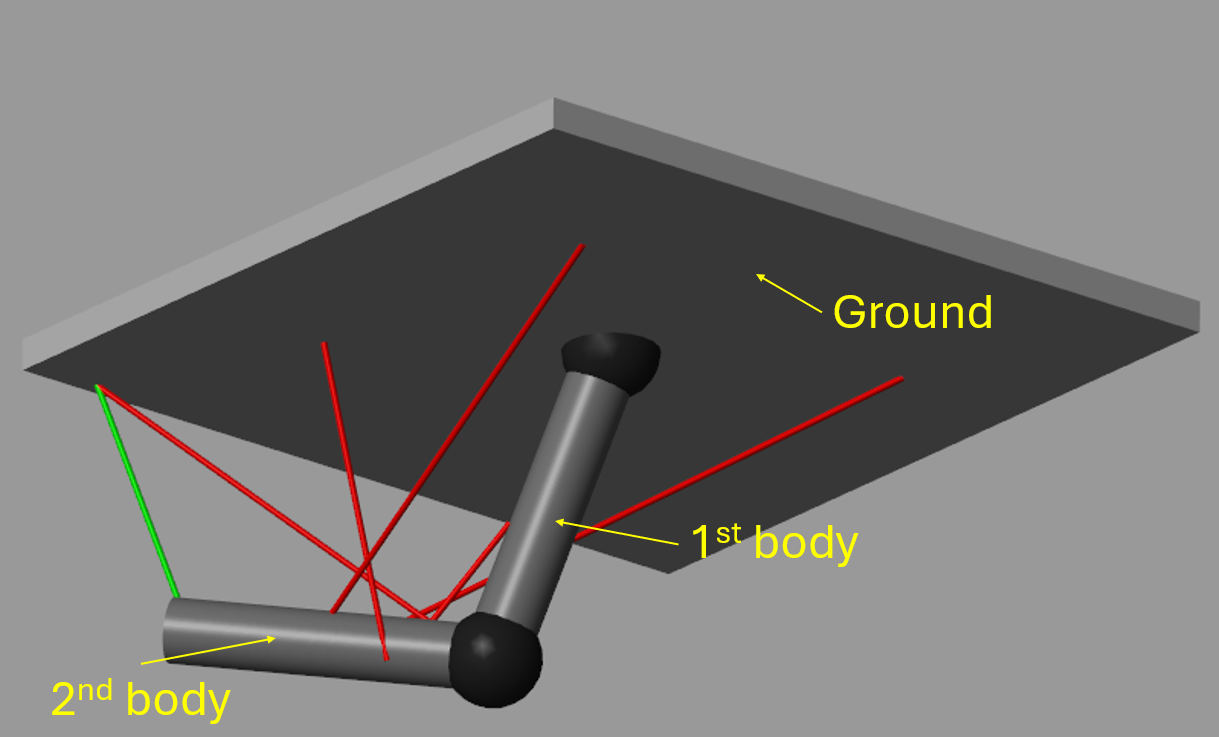}
\caption{Double pendulum system with 6 muscle elements. Forward simulation was performed with various constant forces for each muscle element. Green muscle element was taken as an testing element during the motion for the muscle length Jacobians evaluation.}
\label{fig,double_pend}
\end{figure}

\subsection{Application of the mapping on the biomechanical model of the upper limb}\label{Section, shoulder model}

As an application example, a muscle path approximation of middle serratus anterior element in the Shoulder model by \citet{chadwick_real-time_2014} will be calculated. Experimentally measured motion of humeral abduction in scapular plane with the range of motion from $0^\circ$ to $90^\circ$ was used and random noise to the angles to capture the eventual mismatch of tracked and calculated trajectory was added . This led to a total of 3341 training configurations of the upper limb. This training data was then loaded as a \emph{.mot} file to OpenSim and the muscle lengths and moment arms during the motion were exported. Generated moment arms were mapped to the quaternion coordinates using eq. \ref{eq,moment arms to RS} and \ref{eq, R_Q^* mapping}. The muscle path approximations for both Euler and quaternion model were calculated using the algorithm by \citet{van_den_bogert_implicit_2011}. The stopping criterion of the algorithm for adding more polynomial terms was s.t. the change of the RMSE of an approximation and OpenSim generated values was less than 4\%. The quaternion polynomial approximation was calculated using quaternions and then mapped to equivalent moment arms (using eq. \ref{eq, R_S from R_Q^* mapping} and \ref{eq,Rs to moment arms}) for RMSE calculation and interpretability.

\section{Results}

\subsection{Straight-line muscle in the 3D double pendulum system}
\label{Section, Simple dynamical system}

Mapping the quaternion muscle length Jacobian $\textbf{R}_{Q}^*$ to spatial coordinates using eq. \ref{eq, R_S from R_Q^* mapping} gives the identical $\textbf{R}_S$ as eq. \ref{eq,jacobian in spatial coord from analytical length}. As shown in fig. \ref{fig,moment arms}, mapping $\textbf{R}_S$ to equivalent moment arms using eq. \ref{eq,Rs to moment arms} leads to the identical values as the moment arms calculated analytically using $\partial l_m(\vec{q})/{\partial q_i}$, where $\vec{q}$ is a vector of generalized coordinates \citep{an_tendon_1983}.

\begin{figure}[H]
\centering
\includegraphics[width=1\linewidth]{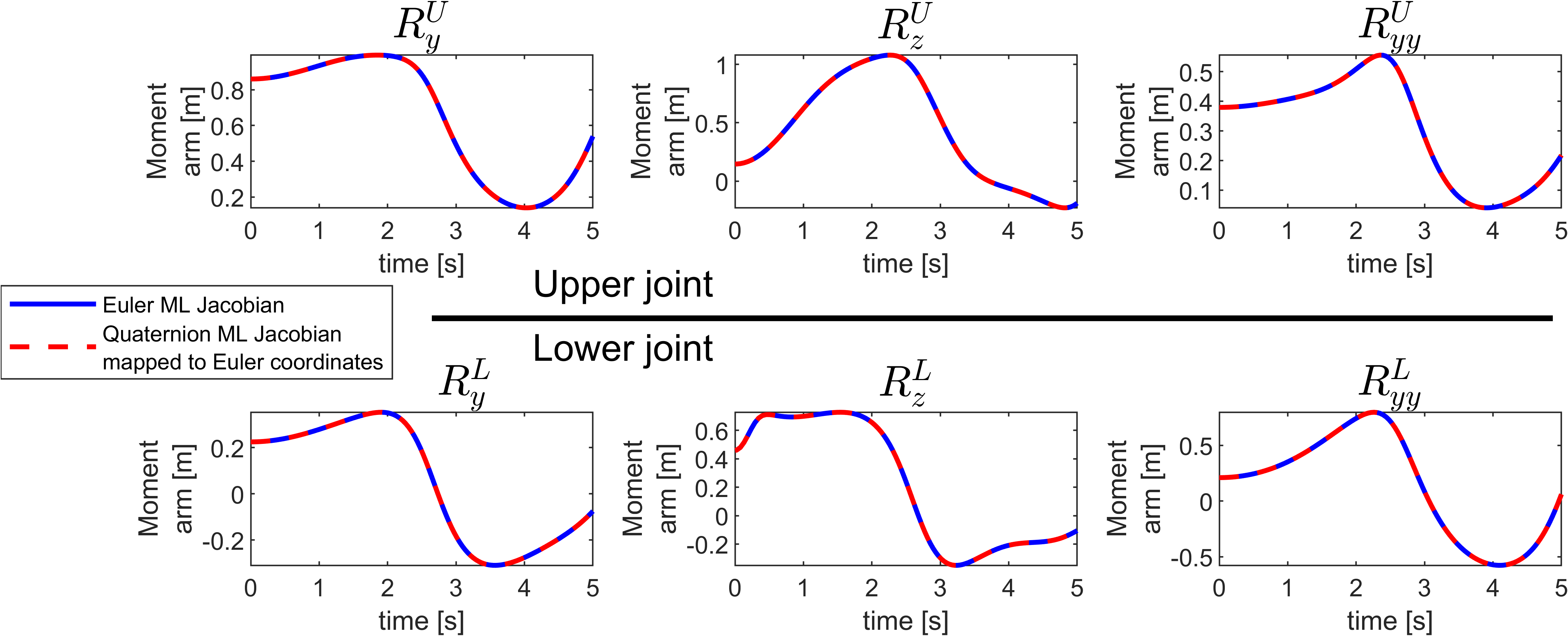}
\caption{Mapping the quaternion muscle length Jacobian to equivalent moment arms (red dashed line) using eq. \ref{eq, R_S from R_Q^* mapping} and \ref{eq,Rs to moment arms} gives us the identical values as the ones computed analytically for YZY sequence (blue line).}
\label{fig,moment arms}
\end{figure}

\subsection{Approximating OpenSim muscle moment arms using quaternions}

The resulting number of polynomial terms of serratus anterior muscle element is 15 for Euler and 13 for quaternion approximation model. The RMSE of exported lengths and polynomial length approximations is $1 mm$ for Euler and $2.36mm$ for quaternion model. The RMSE of exported moment arms and polynomial moment arms approximation is $10.18mm$ for Euler and $10.01mm$ for quaternion model.

\begin{figure}[H]
\centering
\includegraphics[width=1\linewidth]{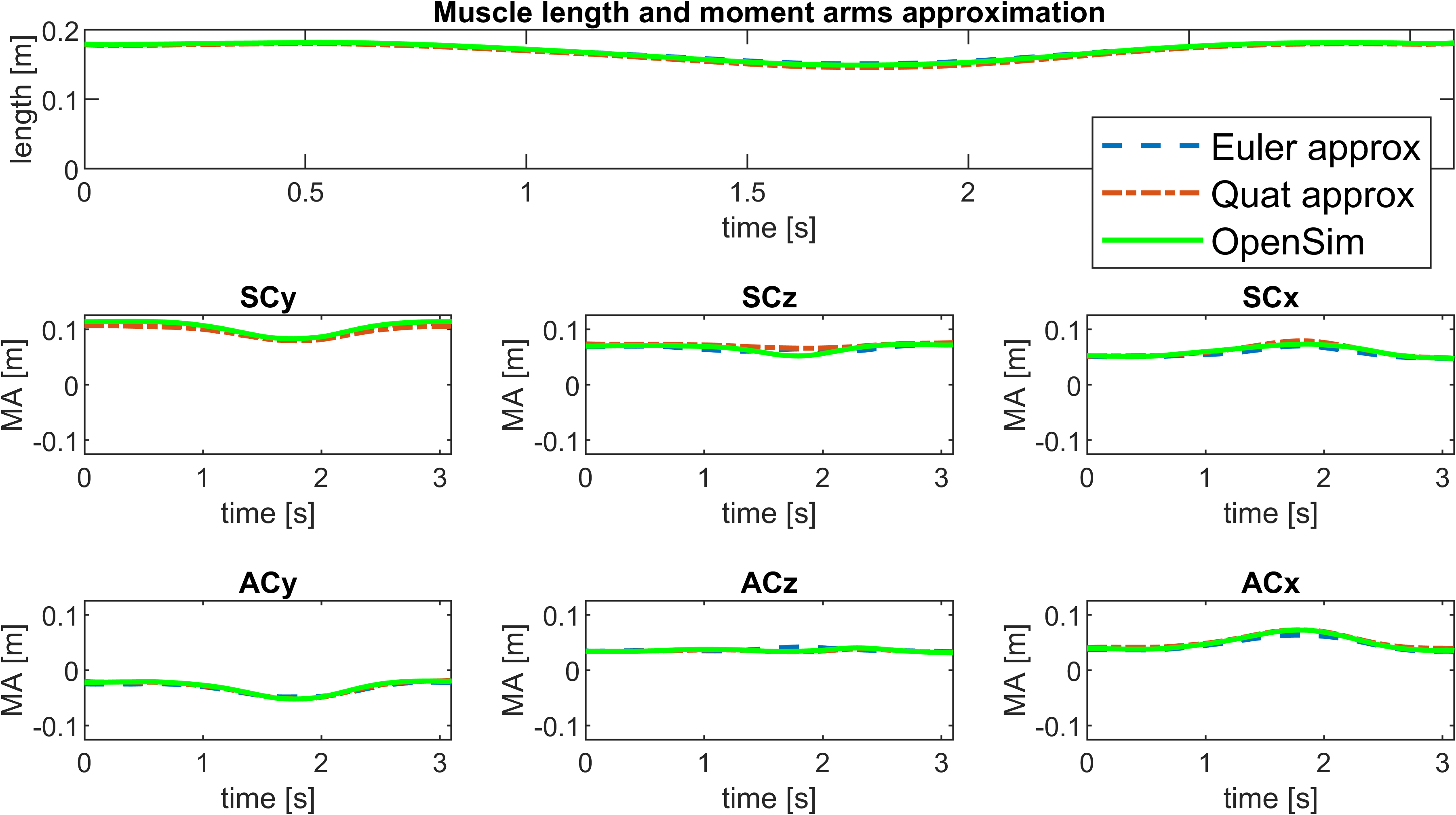}
\caption{Results of approximating a serratus anterior length and moment arms (MA) during scapular abduction. $\textbf{SCy}$, $\textbf{SCz}$ and $\textbf{SCx}$ is the YZX sequence of Euler angles representing sternoclavicular joint and $\textbf{ACy}$, $\textbf{ACz}$ and $\textbf{ACx}$ is the YZX sequence of Euler angles representing acromioclavicular joint. Quaternion muscle length Jacobian was calculated using quaternions and then mapped to Euler coordinates for interpretbility.}
\label{fig, serr_ant_approx}
\end{figure}

\section{Discussion}
\label{Discussion}

In this paper, we presented a mathematical approach to calculate the dynamical contribution from muscle action to a musculoskeletal model described by quaternions. We also introduced a way to map muscle moment arms (in this paper referred to as muscle length Jacobian in Euler coordinates) to quaternion coordinates so that it can be efficiently approximated using polynomial-based algorithms. A simple dynamical system was created to analyze and validate the presented equations.

% exoskeletons cable driven, cable driven robots

For muscles that can be computed as a function of quaternion elements, both approaches without and with constraint (section \ref{Section, Map to external torque Graf} and \ref{Section, mapping with cnst}, respectively) are suitable for muscle torque estimation. Adopting a constraint on the unit quaternion (Eq. \ref{unit quat constraint dt}) lowers computational demands as it reduces the number of derivatives from four to three. However, the constraint-based approach proposed in the current study poses certain limitations. In the eq. \ref{eq,constraint in matrix form} the division by $Q_0$ appears. This means that there is a singularity when axis-angle rotation is exactly $180^\circ$ around an arbitrary axis. We also substituted $Q_0$ for $\sqrt{1-Q_1^2-Q_2^2-Q_3^2}$. Square root only gives positive numbers, that means that the substitution only works when we have a positive $Q_0$. Due to the fact that positive and negative quaternion represents the same rotation, the quaternion can be multiplied by -1 if $Q_0<0$ before evaluating the muscle length Jacobian. In many cases a muscle path approximation is convenient for computational efficiency or for optimal trajectory calculation (smooth differentiable function is needed). If one would like use software like OpenSim for exporting muscle moment arms and map it to quaternion muscle length Jacobian for polynomial approximation algorithms \citep{van_den_bogert_implicit_2011,sobinov_approximating_2020}, only the constraint-based approach is suitable.

In the example of serratus muscle element polynomial approximation there is a difference of $1.36mm$ in the RMSE of the muscle length polynomial approximation between the Euler and quaternion models, while the RMSE of the moment arms is approximately the same. Note that the approximation is performed on different Jacobians, so the resulting polynomial approximation is not exactly the same and depends on the configuration of the algorithm, which is not part of this study.

The proposed method provides robust estimation of muscle torques in dynamic motion that could be particularly suitable for biomechanical systems with potential gimbal lock. Models containing the shoulder joint as one of the most important joints in the human body are obvious candidates for using the proposed framework. The implementation of quaternions in optimal control patient-specific biomechanical simulations may provide better estimation because of the ability to attain all feasible joint trajectories. However, further research is needed to quantify differences between the Euler and quaternion-based musculoskeletal models in specific activities and complex biomechanical models including active properties of muscles. 

\section{Acknowledgments}
This work was supported by the Czech Science Foundation grant number GA23-06920S.

% %% The Appendices part is started with the command \appendix;
% %% appendix sections are then done as normal sections
% \appendix

% \section{Sample Appendix Section}
% \label{sec:sample:appendix}
% Lorem ipsum dolor sit amet, consectetur adipiscing elit, sed do eiusmod tempor section \ref{sec:sample1} incididunt ut labore et dolore magna aliqua. Ut enim ad minim veniam, quis nostrud exercitation ullamco laboris nisi ut aliquip ex ea commodo consequat. Duis aute irure dolor in reprehenderit in voluptate velit esse cillum dolore eu fugiat nulla pariatur. Excepteur sint occaecat cupidatat non proident, sunt in culpa qui officia deserunt mollit anim id est laborum.

% \printbibliography
% \bibliographystyle{elsarticle-harv} 
% \bibliography{bibliography_quat}
\bibliographystyle{unsrtnat}
\bibliography{bibliography_quat}

\end{document}